\documentclass[%
 aip,
 sd,%
 amsmath,amssymb
]{revtex4-1}

\usepackage{graphicx}

\usepackage{amsmath}
\usepackage{amsfonts}

\newcommand{\const}{\operatorname{const}}
\newcommand{\euler}{\operatorname{e}}
\newcommand{\ee}[1]{{\euler}^{#1}}

\newcommand{\ve}{\varepsilon}

\usepackage{bm}
\usepackage{dcolumn}% Align table columns on decimal point

\usepackage{graphicx}
\usepackage{color}
\newcommand{\new}[1]{{{\color{blue}#1}}}

\newcommand{\bcuno}{BC1}
\newcommand{\bcdos}{BC2}
\newcommand{\bctres}{BC3}

\begin{document}
\title{Weakly parametrized Jastrow ansatz for a strongly correlated Bose system}

\author{Yaroslav Lutsyshyn}
\address{Institut f\"{u}r Physik, Universit\"{a}t Rostock, 18059 Rostock, Germany}
\email{yaroslav.lutsyshyn@uni-rostock.de}

%\ead{yaroslav.lutsyshyn@uni-rostock.de}

\begin{abstract}
We consider the Jastrow pair-product wavefunction for strongly correlated Bose systems, in our case liquid helium-4. An ansatz is proposed for the pair factors which consists of a numeric solution to a modified and parametrized pair scattering equation. We consider a number of such simple one-variable parametrizations. Additionally, we allow for a parametrizeable cutoff of the pair factors and for the addition of a long-range phonon tail. This approach results in many-body wavefunctions that have between just one and three variational parameters. Calculation of observables is carried with the Variational Monte Carlo method. We find that such a simple parametrization is sufficient to produce results that are comparable in quality to the best available two-body Jastrow factors for helium. For the two-parameter wavefunction, we find variational energies of $-6.04$~K per particle for a system of one thousand particles. It is also shown that short-range two-body correlations are reproduced in good detail by the two- and three-parameter functions. 
\end{abstract}

\maketitle

\section{Motivation}
Jastrow wavefunction~\cite{Bijl,Jastrow} is a pair-product ansatz for a strongly correlated many-body bosonic ground state. Exactly optimized form of the pair function can be found with the Correlated Basis Functions (CBF) theory and the diagrammatic theory %
\cite{%
Clark1966-MethodOfCorrelatedBasisFunctions,%
Feenberg1959-SimplifiedTreatmentForStrongShortRangeRepulsionsInNParticleSystemsI,%
FeenbergBook1969,%
Krostscheck-in-MBT-V7%
}.
However, such solutions are often either unknown or unavailable from open sources. Meanwhile, the Jastrow wavefunction is widely used as an approximation to the ground state in first-principles quantum many-body methods, both for bosons
\cite{%
Rigol2011-OneDimensionalBosonsFromCondensedMatterSystemsToUltracoldGases,%
Whaley2000-QuantumSolvationAndMolecularRotationsInSuperfluidHeliumClusters,%
Astrakharchik2005-BeyondTheTonksGirardeauGasStronglyCorrelatedRegimeInQuasiOneDimensionalBoseGases,%
Boronat2016-OneDimensional1H2HAnd3H%
}
and as a part of a fermionic wavefunction %
\cite{%
Droghetti2016-DiffusionMonteCarloPerspectiveOnTheSpinStateEnergeticsOfFeNCH,%
Luechow2014-AccurateRotationalBarrierCalculationsWithDiffusionQuantumMonteCarlo,%
Needs2013-QuantumMonteCarloStudyOfTheThreeDimensionalSpinPolarizedHomogeneousElectronGas,%
Maezono2012-ABenchmarkQuantumMonteCarloStudyOfTheGroundStateChromiumDimer,%
Umrigar2012-ApproachingChemicalAccuracyWithQuantumMonteCarlo,%
Astrakharchik2004-EquationOfStateOfAFermiGasInTheBECBCSCrossoverAQuantumMonteCarloStudy,%
Giorgini2011-BCSBECCrossoverAnATwoDimensionalFermiGas%
}. 
In particular, Diffusion Monte Carlo (DMC) requires such an approximation as its guiding (importance-sampling) function; the Path-Integral Ground State Monte Carlo 
 (PIGS) \cite{Schmidt2000-APathIntegralGroundStateMethod}
does not require importance sampling yet benefits greatly when a high-quality approximate ground state is used as its boundary condition 
\cite{%
Galli2009-ExactGroundStateMonteCarloMethodForBosonsWithoutImportanceSampling,%
Boronat2010-HighOrderTimeExpansionPathIntegralGroundState%
}.
%In DMC and PIGS, the approximate nature of the Jastrow function does not impede the computations, as the exact ground state is eventually projected from the trial functions.%
In such cases, the numerical efficiency and convenience of obtaining the trial functions is often important to practitioners. The pair factors of the Jastrow ansatz, when used in the DMC, are often extracted by solving the pair scattering equation for small inter-particle separation, and only to the leading order. Such a solution usually needs to be additionally parametrized with one or several variationally-optimized variables. In this work, we revisit this standard procedure. Instead of parametrizing the leading-order solution to the pair equation, we parametrize the scattering equation itself. The trial functions are optimized in the space of solutions of such parametrized equations. We find that this results in a robust and straight-forward methodology which provides an excellent description of the ground state.

\section{Pair scattering equation and the Jastrow ansatz}
We are considering a system of identical bosons with pairwise interaction potential. The Hamiltonian can be written as
\begin{equation}
H=-\frac{\hbar^2}{2m}\sum_s\nabla_s^2 + \sum_{i<j} V(r_{ij}).
\label{eq:full-Hamiltonian}
\end{equation}
We assume the Bijl-Jastrow ansatz for the many-body ground state of this system,
\begin{equation}
\psi(\bm{r}_1,\dots,\bm{r}_N)=\prod_{i<j}f_2(r_{ij}).
\label{eq:fullpsi}
\end{equation}
The kinetic energy of this function can be written as
\begin{align}
\nabla^2 \psi = & 
\frac12 \sum_s  \bm{\nabla}_s\cdot \left( \bm{\nabla}_s   \prod_{i\ne j}f(r_{ij})  \right)  \\
= &
\frac12 \sum_s\bm{\nabla}_s\cdot \left( \prod_{i\ne j}f(r_{ij})  \right)\left( \sum_{p\ne s}\frac{\bm\nabla_s f(r_{ps})}{f(r_{ps})}  \right)  \\
%= &
%\frac12 \left( \prod_{i\ne j}f(r_{ij})  \right)
%\sum_s\left( \sum_{p\ne s}\frac{\nabla_s^2 f(r_{ps})}{f(r_{ps})}  \right) \\
%&+
%\frac12 \left( \prod_{i\ne j}f(r_{ij})  \right)
%\sum_s
%\left( \sum_{t\ne s}\frac{\bm\nabla_s f(r_{ts})}{f(r_{ts})}  \right)
%\cdot\left( \sum_{p\ne s}\frac{\bm\nabla_s f(r_{ps})}{f(r_{ps})}  \right) \\
%& -
%\frac12 \left( \prod_{i\ne j}f(r_{ij})  \right)
%\sum_s\left( \sum_{p\ne s}\frac{ | \bm\nabla_s^2 f(r_{ps}) | ^2}{f^2(r_{ps})}  %\right) \\
= &
\frac12 \left( \prod_{i\ne j}f(r_{ij})  \right)
\sum_s\left( \sum_{p\ne s}\frac{\nabla_s^2 f(r_{ps})}{f(r_{ps})}  \right) \nonumber \\
&+
\frac12 \left( \prod_{i\ne j}f(r_{ij})  \right) \times \nonumber \\
\times&
 \sum_{\substack{t,p,s  \\ t,p \ne s \\ t \ne p }}\frac{  f'(r_{ts}) f'(r_{ps})}{f(r_{ts}) f(r_{ps})} 
\frac{ (\bm{r}_t-\bm{r}_s)\cdot(\bm{r}_p-\bm{r}_s)  }{r_{ts}r_{ps}}  .
\end{align}
In a homogeneous system, the second term vanishes
if one assumes that the three-body correlations are sufficiently weak,
and the eigenvalue equation of Hamiltonian (\ref{eq:full-Hamiltonian}) is fulfilled so long as factors $f_2$ satisfy the pair equation
\begin{equation}
-\frac{\hbar^2}{m}\nabla^2 f_2(r) + V(r) f_2(r) = \varepsilon f_2(r).
\label{eq:pair-equation-f}
\end{equation}
The energy $\varepsilon$ is equal to $\sim{E_0}/N^2$, $E_0$ being the eigenenergy of the many-body ground state, and must in fact vanish for an infinite system.
We retain this term as our calculations use finite periodic systems and $\varepsilon$ can be used to satisfy corresponding boundary requirements.
Direct solutions of the pair scattering equation similar to (\ref{eq:pair-equation-f}) were used by \new{Pandharipande}
\cite{
	Pandharipande1972-VariationalCalculationOfNuclearMatter,%
	Pandharipande1971-HyperonicMatter,%
	Pandharipande1971-DenseNeutronMatterWithRealisticInteractions%
}
as part of a variational ansatz for nuclear matter. However, as we will see below, direct solutions of this equation are far from optimal for our systems of interest.

It was realized early on that the dominant contribution 
to the quality of the pair factor comes from its behavior at small distances, 
where the potential energy diverges for a hard-core interaction.
Indeed, to this day solving Eq.~(\ref{eq:pair-equation-f})
to the first leading order at short distances 
remains the standard prescription when a simple, few-parameters
function is desired.
This is often understood as solving the Kato cusp condition 
\cite{Kato1957-OnTheEigenfunctionsOfManyParticleSystemsInQuantumMechanics}
for the divergence in $V(r)$, similar to the way one handles interactions such
as \emph{e.g.} electronic states in the Coulomb potential or states in 
a finite-strength contact potential. Such terminology is not strictly correct for the potentials with an analytical hard-core part. In fact, at short distances, the optimal pair factors for hard-core interactions vanish exponentially faster than the divergence of the corresponding potential or kinetic energy contributions. The expectation values for both the potential and kinetic energies exist and carry no irregularities. Likewise, the wavefunction has well-defined derivatives everywhere where it does not vanish. This is the reason why the leading-order solution of Eq.~(\ref{eq:pair-equation-f}) can be additionally parametrized without forfeiting the variational principle.
On the contrary, systems with a true cusp condition must satisfy their cusp condition exactly in order to cancel non-analytic terms that arise in the Schr\"odinger equation; such pair factors cannot be meaningfully parametrized in the vicinity of the divergence. 

We write the pair factors in a positive-defined form as
\begin{equation}
f_2=\ee{u_2},
\end{equation}
which is the conventional pseudopotential form
but with the prefactor $1/2$ omitted for the sake of 
simplicity.
The pair equation (\ref{eq:pair-equation-f}) reads
\begin{equation}
-\frac{\hbar^2}{m}\left( u_2''(r)+\frac{D-1}{r}u_2'(r) + [u'_2(r)]^2 \right)
+ V(r) = \varepsilon,
\label{eq:pair-equation-u-start}
\end{equation}
where $D$ is the physical dimensionality of the system.
If the pair potential energy diverges at small distances as
\begin{equation}
V(r)=\upsilon r^{-k}
\end{equation}
with $k>2$,
the other leading term in Eq.~(\ref{eq:pair-equation-u-start})
is in fact $(u'_2)^2$ and one has WKB-like 
solution
\begin{equation}
u'_2(r)=+\sqrt{\frac{m\upsilon}{\hbar^2} r^{-k}},
\end{equation}
and
\begin{equation}
u_2(r)=-\frac{2}{k-2}\sqrt{\frac{m\upsilon}{\hbar^2}} r^{-\frac{k}{2}+1} + \const.
\end{equation}
In practice, this solution does not yield an optimal energy
but is in fact significantly improved by additional parametrization,
\begin{equation}
u_2(r)=-\alpha {r}^{-\frac{k}{2}+1} + \const,
\label{eq:leading-term-u2}
\end{equation}
where the constant $\alpha$ can be determined by minimizing the variational energy.
As mentioned above, such a parametrization is possible because every term of the pair equation (\ref{eq:pair-equation-f}) remains analytical, and in fact even infinitely differentiable.
Indeed,
\begin{equation}
\nabla^2 f_2 \sim
V(r)f_2(r)\sim\frac{1}{r^k}\exp\left(-\frac{\alpha}{r^{k/2-1}}\right)\rightarrow 0.
\end{equation}
Moreover, the power of the divergence in $u_2$ can also be treated 
variationally for the same reason. 
The above approach alone allows to account for most of the correlation energy
of the dense strongly correlated bosonic systems. 

\section{Improvements beyond the leading order}
Numerous works have been devoted to improving the pair ansatz beyond the form of Eq.~(\ref{eq:leading-term-u2}). One of the most studied
systems in this regard is liquid helium-4. This system has been the subject of about every quantum Monte Carlo method that is applicable, which included a large amount of work with explicit Variational Monte Carlo with the Jastrow ansatz. The simplest variational ansatz for helium consists of the parametrized leading-term solution proposed by McMillan and by Schiff and Verlet
\cite{%
	McMillan,%
	SchiffVerlet-1967%
}, or similar solutions 
\cite{%
	LevesqueSchiff1971-Fluid-SolidPhaseTransitionOfAHardSphereBoseSystem,%
	KalosLevesqueVerlet1974-HeliumAtZeroTemperatureWithHardSphereAndOtherForces%
}.
The improvements beyond the leading order for helium 
can be seen as falling into the following categories: 
(a) addition to the leading term (\ref{eq:leading-term-u2}) of various explicit short- and long-range corrections derived from known general properties of the system
\cite{%
	ReattoChester1966-TheGroundStateOfLiquidHeliumFour,%
	ReattoChester1967-PhononsAndThePropertiesOfABoseSystem,%
	Henderson1979-OnWavefunctionsForTheGroundStateOfLiquid4He,%
	Reatto1980-JastrowWaveFunctionAndCorrelationsOfTheLennardJonesBoseFluid%
},
(b) parametrization of the mid-range behavior of $u_2$ focused primarily on improving the 
variational energy or structural properties 
\cite{%
	Reatto1974-HowGoodCanJastrowWavefunctionsBeForLiquidHeliumFour,%
	Graben1978-OnJastrowWavefunctionsContainingAttractiveCorrelations,%
	Reatto1980-JastrowWaveFunctionAndCorrelationsOfTheLennardJonesBoseFluid,%
	ChinUnpublishedPairFunction%	
}, and 
(c) extracting the solution to
the pair equation (\ref{eq:pair-equation-u-start}) beyond the leading order, 
or using such solutions as a basis set 
\cite{%
	Vitiello1992-OptimizationOfHeWaveFunctionsForTheLiquidAndSolidPhases,%
	Vitiello1999-VariationalMethodsForHeUsingAModernHeHePotential,%
	Moroni1995-EulerMonteCarloCalculationsForLiquidHe4AndHe3%
}.
These approaches are considered in some detail below. Additionally, one can use the pair functions obtained with Paired Phonon Analysis (PPA) method 
\cite{%
	Campbell1978-FunctionalOptimizationOfTheJastrowWaveFunctionForLiquid4He, %
	CampbellFeenberg1969-PairedPhononAnalysisForTheGroundStateAndLowExcitedStatesOfLiquidHelium%
}. 
For instance, PPA optimization of Campbell and Pinski was used as a guiding function in Ref.~%
\onlinecite{Chester1981-ModernPotentialsAndThePropertiesOfCondensedHeFour}.
This method requires the HNC approximation and is out of our scope. 
The best forms of the Jastrow function miss the ground state of liquid helium by about one degree Kelvin per particle. A large part of this energy can be accounted for by including three-body correlations into the wavefunction 
\cite{%
	Feenberg1974-GroundStateOfAnInteractingBosonSystem,%
	Schmidt1980-VariationalMonteCarloCalculationsOfLiquid4HeWithThreeBodyCorrelations,%
    Berdahl1974-Three-particleCorrelationsInTheGroundStateOfABoseFluid,%
	Vitiello1992-OptimizationOfHeWaveFunctionsForTheLiquidAndSolidPhases,%
	Moroni1995-EulerMonteCarloCalculationsForLiquidHe4AndHe3,%
	Vitiello2003-AnalysisOfTheContributionsOfThreeBodyPotentialsInTheEquationOfStateOfHe4%
}.
Partially, the three-body correlations can be captured by the shadow wavefunction (SWF) method 
\cite{
	Vitiello1988-VariationalCalculationsForSolidAndLiquidHe4WithAShadowWaveFunction,%
	Reatto1988-ShadowWaveFunctionForManyBosonSystems%
} which has since taken over explicit wavefunctions for the variational description of liquid helium.
However, our interest is limited to the explicit Jastrow pair terms.

(a) This approach relies on knowing the general properties of the static structure function of the system. It allows to recreate the physically important features that otherwise are missed by the leading-order solution. Most remarkable of these corrections is the infinite-range tail in $u_2$ which must arise in the presence of zero-point phonons in the system
\cite{ReattoChester1967-PhononsAndThePropertiesOfABoseSystem}. 
Other corrections to $u_2$ were initially tied to the mid-range excitation structure of helium, that is, to the presence of the roton minimum 
\cite{Reatto1978-JastrowWaveFunctionForCondensedPhasesOfBoseParticlesHardSphereSystem}. 
As the mid-range behavior is critical for achieving good varitational energy, parametrising $u_2$ in this region has a significant impact on the variational energy, disregarding the origin of such correction. Overall, this approach leads to a rapid escalation of the parametrization of the pair factor. The leading order term must still be present, contributing one or two parameters. The phonon tail adds between one and two parameters (speed of sound,
and a distance cutoff $k_c^{-1}$), and at least three parameters are required for a simple Gaussian mid-range correction (its location, width and strength).

(b) If one is solely interested in minimizing the variational energy, or in replicating a known observable such as the pair distribution function, it is possible to this end to employ heavy parametrization of the pair function 
\cite{%
	Reatto1974-HowGoodCanJastrowWavefunctionsBeForLiquidHeliumFour,%
	ChinUnpublishedPairFunction%
}.
Such attempts have been quite successful, but in all cases relied on on using six and more parameters. Here we are seeking approaches that are more straight-forward methodologically.

(c)  Trying to directly solve the pair scattering equations leads to 
the pair functions that are extremely poor in accounting for the correlation 
energy. Moreover, as will be shown below, such energies depend very strongly 
on the size of the simulation cell.
Pandharipande and Schmidt used the solutions of the pair equation as a variational anzatz for helium
\cite{Pandharipande1977-VariationalCalculationsOfSimpleBoseSystems}. 
A sharp cutoff allowed to limit the energy divergence, and the offset $\lambda$ in Ref.\ 
\onlinecite{Pandharipande1977-VariationalCalculationsOfSimpleBoseSystems}, used to satisfy boundary condition for $f_2$ at the cutoff,
is equivalent to $\varepsilon$ in (\ref{eq:pair-equation-u-start}).
Unfortunately, the energy in Ref.\ 
\onlinecite{Pandharipande1977-VariationalCalculationsOfSimpleBoseSystems} was extracted in an approximate way.
A more productive approach, with excellent variational results, was reported by Vitiello and Schmidt in Ref.~\onlinecite{Vitiello1992-OptimizationOfHeWaveFunctionsForTheLiquidAndSolidPhases}.
It was shown that when the pair equation is treated as an eignevalue 
problem, the set of lowest eigenfunctions can be used as a basis 
for expressing the pair function. 
That is, their approach uses the pair equation as a \emph{source} of good variational functions.
On the other hand, the need to combine multiple eigenstates
means one has once again to deal with as many as ten variational parameters.
The present work can be seen as extending the approach of Pandharipande and Schmidt. However, instead of parametrizing a linear combination of functions arising from the pair equation, we choose to parametrize the equation itself. The pair function is selected from the space of solutions of a modified and parametrized version of the differential Eq.~(\ref{eq:pair-equation-u-start}). 

\section{Method}

\subsection{Modified pair equations}

\begin{figure}[t]
\begin{centering}
\includegraphics[width=0.5\textwidth]{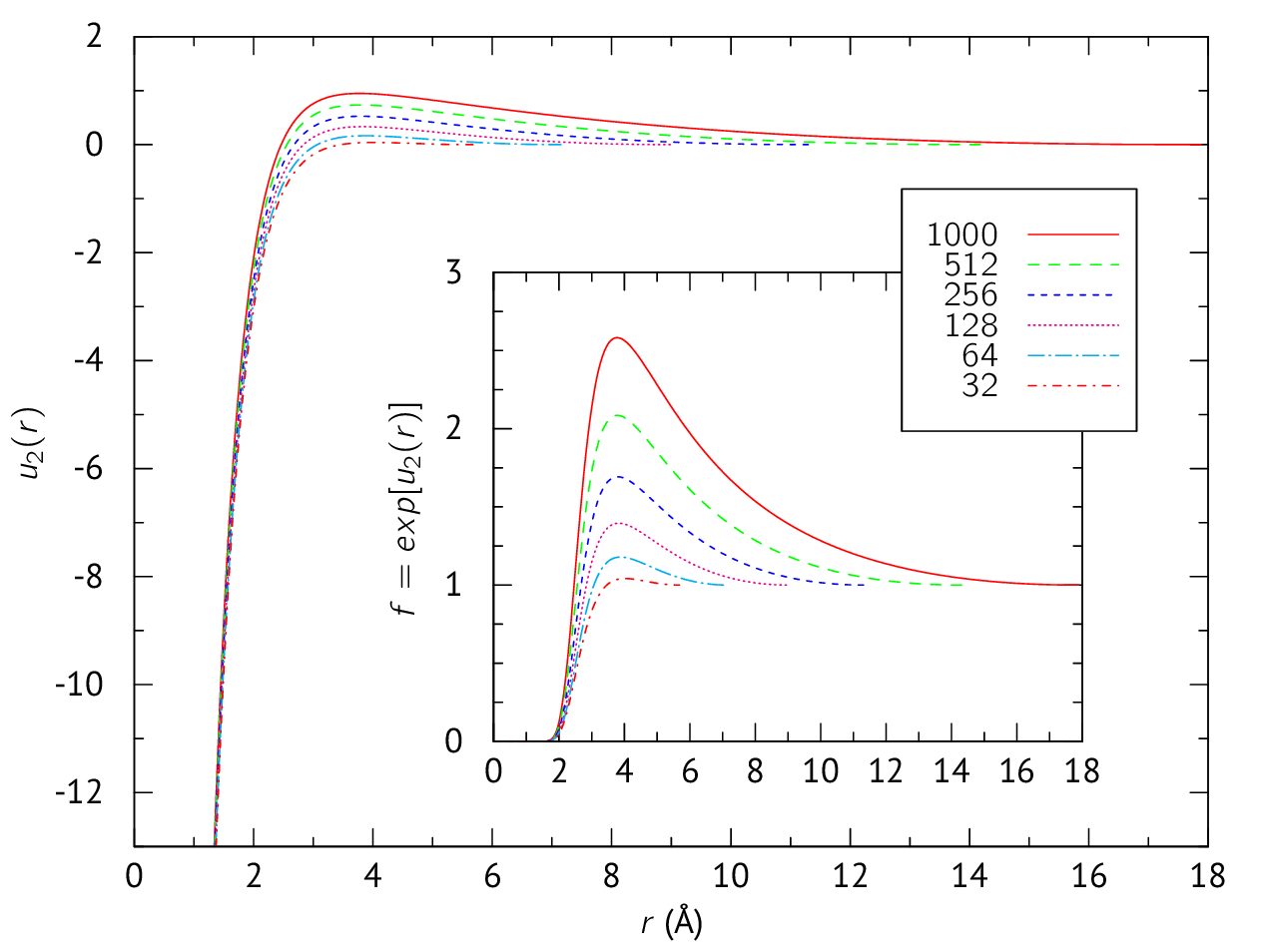}
\caption{
Solutions of the unaltered pair scattering Eq.~(\ref{eq:pair-equation-u-start}) for varying cutoff lengths. The equation is solved for helium with the Aziz HFD-B(HE) \cite{AzizII} interparticle potential. The cutoff lengths correspond to cubic simulation cell with helium at its equilibrium density containing 32, 64, 128, 256, 512, and 1000 atoms. While the solutions coincide at small interparticle separations, the mid-range maximum increases with system size. The inset shows the pair factor $f_2=\exp[u_2(r)]$. %
\label{fig:u2straight}%
}
\end{centering}
\end{figure}

Straight-out solution of the pair Eq.~(\ref{eq:pair-equation-u-start})  for any potential with an attractive part of the potential contains an exaggerated feature at mid-range separations. The larger is the simulation cell, the further is the solution cutoff, and the stronger becomes this peak. Example  solutions for helium are shown in Fig.~(\ref{fig:u2straight}). We observed that this feature is especially sensitive to the presence of the second term on the l.h.s. of~Eq.~(\ref{eq:pair-equation-u-start}). Thus we considered several parametrized modifications of that term.

With substitution 
\begin{equation}
v_2=u'_2,
\label{eq:v2def}
\end{equation}
and additional notation
\begin{equation}
g_\ve(r)=\frac{V(r)-\ve}{\hbar^2/m},
\label{eq:def-g}
\end{equation}
the pair equation reads 
\begin{equation}
-v'_2 -\frac{D-1}{r} v_2 - (v_2)^2 + g_\ve=0,
\label{eq:unmodified-v2}
\end{equation}
which is nonlinear but can be solved numerically. The energy parameter $\ve$ is used to satisfy the boundary conditions.

Below we list the parametrized equations which we consider. Each equation contains one variational parameter, but may also be modified in that the ``dimensionality'' term is either switched in sign or removed altogether. For clarity of notation, each equation uses a uniquely named variable for its parametrization.
\begin{itemize}
\item[(A)]
``Effective mass'' modifications,
\begin{align}
-v'_2 -\frac{D-1}{r} v_2 - (v_2)^2 + {\mu_1} g_\ve &=0  \label{eq:original-3}, \\
-v'_2 + \frac{D-1}{r} v_2 - (v_2)^2 + {\mu_2} g_\ve &=0 \label{eq:original-4}, \\
-v'_2  - (v_2)^2 + {\mu_3} g_\ve &=0                    \label{eq:original-5}.
\end{align}
\item[(B)]
Prefactor to the quadratic term,
\begin{align}
-v'_2 -\frac{D-1}{r} v_2 - k_1 (v_2)^2 + g_\ve &=0 \label{eq:original-6}, \\
-v'_2 +\frac{D-1}{r} v_2 - k_2 (v_2)^2 + g_\ve &=0 \label{eq:original-7}, \\
-v'_2 - k_3 (v_2)^2 + g_\ve &=0                    \label{eq:original-8}. 
\end{align}
\item[(C)]
Prefactor to the linear (``dimensionality'') term in $v_2$,
\begin{align}
-v'_2 -\frac{\delta-1}{r} v_2 - (v_2)^2 + g_\ve &=0 \label{eq:original-9}.
\end{align}
%\item[(D)]
%Distance power modification to the linear term,
%\begin{align}
%-v'_2 -\frac{D-1}{r^{\nu_1}} v_2 - (v_2)^2 + g_\ve  &=0 \label{eq:original-10},\\
%-v'_2 -\frac{D-1}{-r^{\nu_2}} v_2 - (v_2)^2 + g_\ve &=0 \label{eq:original-11}.
%\end{align} 
\item[(D)]
Suppression of the divergence of the linear term at short distance,
\begin{align}
-v'_2 -\frac{D-1}{r+s_1} v_2 - (v_2)^2 + g_\ve &=0 \label{eq:original-12},  \\
-v'_2 +\frac{D-1}{r+s_2} v_2 - (v_2)^2 + g_\ve &=0 \label{eq:original-13}.
\end{align}
\item[(E)]
Suppression of the linear term at large distances,
 \begin{align}
-v'_2 -\frac{D-1}{r/[1-(r/L_c)^{1/m_1}]} v_2 - (v_2)^2 + g_\ve &=0 \label{eq:original-14}, \\
-v'_2 +\frac{D-1}{r/[1-(r/L_c)^{1/m_2}]} v_2 - (v_2)^2 + g_\ve &=0 \label{eq:original-15}.
\end{align}
The terms are suppressed when approaching the cutoff distance $L_c$.
\item[(F)]
Finally, a prefactor in the first term,
\begin{align}
-c_1 v'_2 -\frac{D-1}{r} v_2 - (v_2)^2 + g_\ve &= 0 \label{eq:original-16}, \\  
-c_2 v'_2 +\frac{D-1}{r} v_2 - (v_2)^2 + g_\ve &= 0 \label{eq:original-17}, \\
-c_3 v'_2 - (v_2)^2 + g_\ve &=0                     \label{eq:original-18}.
\end{align}
\end{itemize}

\subsection{Boundary conditions, cutoffs and long-range behavior \label{sec:boundary-conditions}}

%%%%%%%%%%%%%%%%%%%%%%%%%%%%%%%%%%%%%%%%%%%%%%%%%%%%%%%%%%%%%%%%%%%%%%%%%%%%%%%%%%%%%%%%%%%%%%%%%%%%%%%%%%%%%%
%%%%%%%%%%%%%%%%%%%%%%%%%%%%%%%%%%%%%%%%%%%%%%%%%%%%
%%%%%%%%%%%%%%%%%%%%%%%%%
%%%%%%%%%%%%%

\begin{table*}
\begin{ruledtabular}
\caption{ \label{tab:results-helium-bc1}
Optimized variational energy for the helium system with 1024 atoms at density $21.86$~nm\textsuperscript{3}.
The wavefunctions are one-, two-, and three-parameter functions defined by the boundary conditions specified in Section~\ref{sec:boundary-conditions}, with pair factors given by the modified pair Eqs.~(\ref{eq:original-3}--\ref{eq:original-18}). 
For the BC2 and BC3 functions, solution cutoff $L_c$ optimized to the same value for all equations, as shown below.
Parameters $s_1$ and $s_2$ were tested in the range $0<s_{1,2}<25$~\AA.
}
\begin{tabular}{ll|ll|ll|lll}
%Pair factor         &&  { BC1 }     &  $E/N$~(K)   &   { BC2 } &  $E/N$~(K) & BC3  &  $E/N$~(K) \\ %       \\  $L_c=1.6\sigma$  \\ BC3: Lc   1.55 
%                    && \multicolumn{2}{l|}{     }  &   \multicolumn{2}{l|}{ ($L_c=4.09$~\AA)  }  & \multicolumn{2}{l}{ ($L_c=3.96$~\AA,~$c=0.253$) }     \\
Pair factor         &&  { BC1 }           &                &   \multicolumn{2}{l|}{  BC2 ($L_c=4.09$~\AA)  }   &  \multicolumn{3}{l}{ BC3 ($L_c=3.96$~\AA) }    \\ %       \\  $L_c=1.6\sigma$  \\ BC3: Lc   1.55 
                    &&  param.            &   $E/N$~(K)    &   param.    &                        $E/N$~(K)    &  param.         &  $c$  &     $E/N$~(K)  \\
%
%             & Parameter &  $E/N$~(K)  &  \\  
\hline
Eq.~(\ref{eq:original-3})     & $\mu_1$  &  ---     &             &         &                &&\\
Eq.~(\ref{eq:original-4})     & $\mu_2$  &  $1.2$   & $-5.03(1)$  &         &                &&\\
Eq.~(\ref{eq:original-5})     & $\mu_3$  &  $0.98$  & $-5.86(1)$  & $1.00$  & $-6.04(1)$     &&\\
Eq.~(\ref{eq:original-6})     & $k_1$    &  ---     &             &         &                &&\\
Eq.~(\ref{eq:original-7})     & $k_2$    &  $1.3$   & $-5.40(1)$  &         &                &&\\
Eq.~(\ref{eq:original-8})     & $k_3$    &  $0.98$  & $-5.87(1)$  & $1.00$  & $-6.04(1)$     &&\\
Eq.~(\ref{eq:original-9})     & $\delta$ &  $0.90$  & $-5.87(1)$  & $1.12$  & $-6.04(1)$     &  $1.02 $  & $0.25$ & $-6.06(1)$ \\
%Eq.~(\ref{eq:original-10})    & $\nu_1$  &  ---     &             &         &                &&\\
%Eq.~(\ref{eq:original-11})    & $\nu_2$  &  $4$     & $-5.185(2)$ &         &                &&\\
Eq.~(\ref{eq:original-12})    & $s_1$    &  ---     &             &         &                &&\\
Eq.~(\ref{eq:original-13})    & $s_2$    &  $20$~\AA& $-5.84(1)$  &         &                &&\\
Eq.~(\ref{eq:original-14})    & $m_1$    &  ---     &             &         &                &&\\
Eq.~(\ref{eq:original-15})    & $m_2$    &  $8$     & $-5.80(1)$  &         &                &&\\
Eq.~(\ref{eq:original-16})    & $c_1$    &  $1.36$  & $-5.85(1)$  & $1.38$  & $-6.03(1)$     &  $1.40 $  &  $0.26$  &$-6.06(1)$ \\
Eq.~(\ref{eq:original-17})    & $c_2$    &  $0.66$  & $-5.88(1)$  &         &                &&\\
Eq.~(\ref{eq:original-18})    & $c_3$    &  $1.02$  & $-5.87(1)$  & $1.00$  & $-6.04(1)$     &&\\
%McMl [19xx]               & $b$         & 3.07~\AA &  $-5.725$ [non-corr] \\
\end{tabular}
\end{ruledtabular}
\end{table*}

In the above form, Eqs.~(\ref{eq:original-3}--\ref{eq:original-18}) in combination with boundary conditions define families of one-parameter variational functions. For clarity, we refer to solutions of a particular equation by its modification variable. For example, $v_2(r,s2)$ is the solution of Eq.~(\ref{eq:original-13}). We use the notation $v_2(r,\cdot)$ to refer to the solution of either of the modified pair equations.

The Monte Carlo calculation uses periodic boundary conditions with the nearest-neighbor convention. Thus the many-body wavefunction  is periodic by construction. To avoid cusps and corresponding energy corrections, in all cases we demand that the pair factor has zero derivative at the periodicity cutoff boundary $L_b$ (half the smallest simulation cell dimension),
\begin{equation}
v_2(L_b)=0.
\label{eq:boundary-box}
\end{equation}
To formally define the solutions of the Eq.~(\ref{eq:v2def}), we also demand
\begin{equation}
u_2(L_b)=0,
\label{eq:boundary-box-u2}
\end{equation}
which can always be satisfied.

We consider three families of boundary conditions, which result in wavefunctions with one, two, or three variational parameters. 
%\begin{itemize}
%\item

(\bcuno)
Pair factors consist entirely of the solution of one of the Eqs.~(\ref{eq:original-3}--\ref{eq:original-18}). Boundary condition (\ref{eq:boundary-box}) is satisfied by numerically solving for the value of the parameter $\ve$ (see Eq.~\ref{eq:def-g}). This defines a one-parameter Jastrow function.
%\item

(\bcdos)
Modified pair equations are solved up to a cutoff distance $L_c \le L_b$.
Parameter $\ve$ is found numerically to satisfy
\begin{equation}
v_2(L_c,\cdot)=0.
\end{equation}
The pair factor is defined piecewise,
\begin{equation}
v_2(r)=
\begin{cases}
v_2(r,\cdot), & r\le L_c, \\
0,            & L_c < r \le L_b.
\end{cases}
\label{eq:bc2}
\end{equation}
We treat the cutoff distance $L_c$ as a variational parameter, 
which results in a two-parameter variational function.  

%\item
(\bctres)
We define the phonon tail \cite{ReattoChester1967-PhononsAndThePropertiesOfABoseSystem}
\begin{equation}
u_2^{ph}=-\frac{c}{r^2}-\frac{c}{(2L_b-r)^2}+\frac{2c}{L_b^2},
\end{equation}
and correspondingly
\begin{equation}
v_2^{ph} = \frac{2c}{r^3}-\frac{2c}{(2L_b-r)^3}.
\end{equation} 
These functions satisfy Eqs.~(\ref{eq:v2def}),(\ref{eq:boundary-box}),(\ref{eq:boundary-box-u2}).
Modified pair equations are solved up to a cutoff distance $L_c \le L_b$,
like in the previous case. At $r=L_c$, the pair solution is matched to the phonon tail,
\begin{equation}
v_2(L_c,\cdot)=v_2^{ph}(L_c),
\end{equation}
by numerically solving for $\ve$.
The pair factor is once again defined piecewise,
\begin{equation} 
v_2(r)=   
\begin{cases}  
v_2(r,\cdot), & r\le L_c, \\ 
v_2^{ph}(r),            & L_c < r \le L_b. 
\end{cases} 
\label{eq:bc3}
\end{equation}
Prefactor $c$ incorporates the speed of sound and is treated as a variational parameter. Thus this wavefunction has three parameters.
%\end{itemize}

\section{Results}

We carry the calculations with Variational Monte Carlo. The Hamiltonian (\ref{eq:full-Hamiltonian}) is employed with a widely accepted Aziz HFD-B(HE) pair potential, also known as Aziz-II \cite{AzizII}. Size of the cell is selected to provide particle density of 21.86~nm\textsuperscript{3}, corresponding to the equilibrium density of helium at its saturated vapor pressure. We used 1024 atoms in a periodic cubic cell, allowing for the cutoff at $L_b=18.0$~\AA.%, and the potential energy correction due to the finite size of the system of just $-0.080$~K per particle. 
The computation was carried with the QL package on an array of graphical accelerators \cite{Lutsyshyn2015-FastQuantumMonteCarloOnAGPU}.

\subsection{One-parameter wavefunctions}

Summary of the results for the one-parameter wavefunction with \bcuno\ boundary conditions are shown in Table~\ref{tab:results-helium-bc1}.
A number of equations either failed to produce a pair function with a bound state, or did not allow for a value of $\ve$ which would satisfy the boundary conditions. All of these are the equations with the unmodified second, or ``dimensionality'', term on the l.h.s. term of Eq.~(\ref{eq:unmodified-v2}). Only Eq.~(\ref{eq:original-16}) could be optimized with this term fully intact. On the other hand, all of the solution families that could be optimized did in fact result in rather satisfactory energies. This can be attributed to the fact that the close-range behavior \new{is} captured properly by all of these functions.

\begin{figure*}[t]
\begin{centering}
\includegraphics[width=\textwidth]{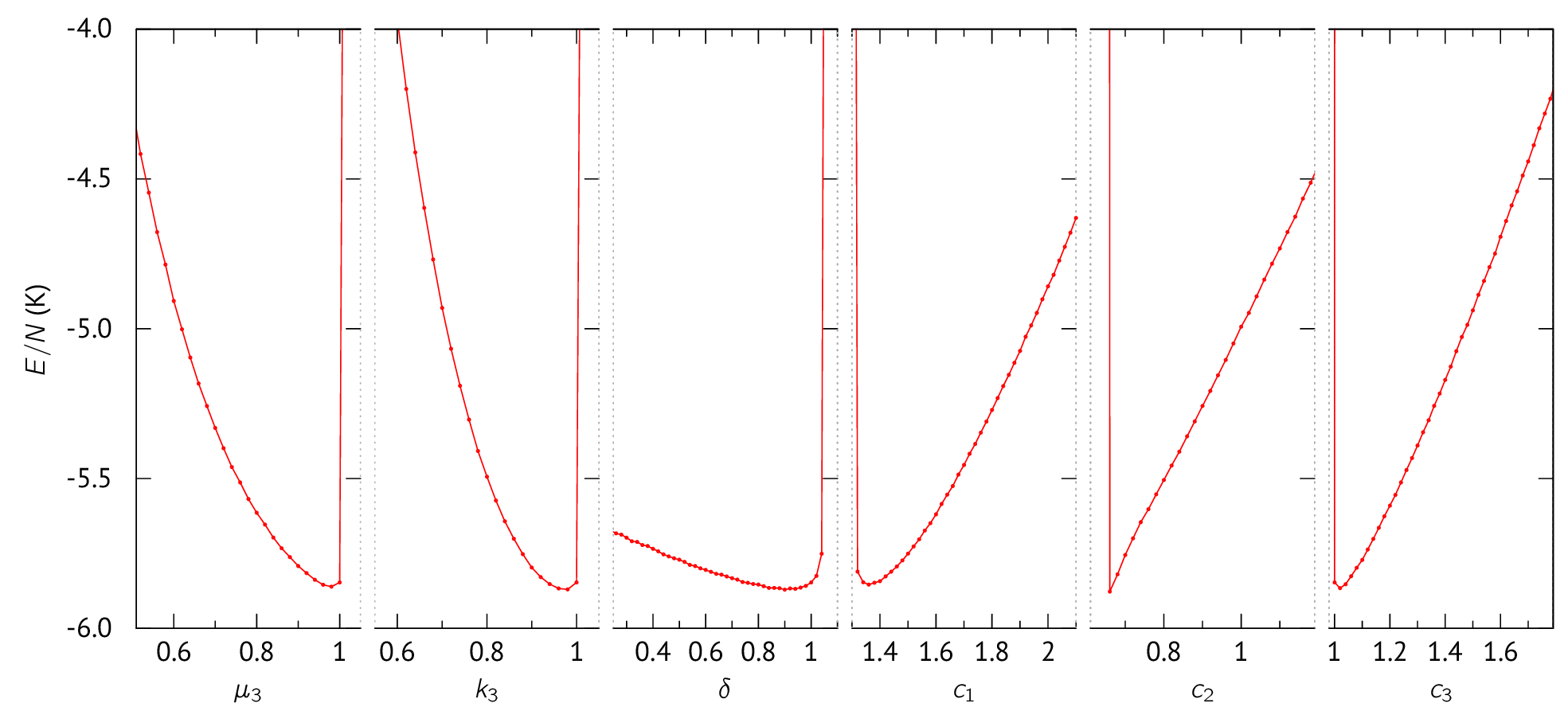}
\caption{
\label{fig:bc1-zoom}
Detail of the variational energy dependence near the minima for the one-parameter BC1 wavefunctions. The panel shows the equations with lowest variational energy. Parameters are labeled corresponding to the notation of Eqs.~(\ref{eq:original-3}--\ref{eq:original-18}). The dependence of energy for the solutions of Eq.~(\ref{eq:original-17}) with coefficient $c_2$ appears to be non-analytic near the minimum.
}
\end{centering}
\end{figure*}

Parameters $\mu_3$, $k_3$, and $c_3$ optimized to a value close to unity. Exactly at unity, these equation all reduce to the same expression,
\begin{equation}
-v_2'-(v_2)^2+g_\ve=0.
\label{eq:original-chopped}
\end{equation}
This is also the limit for large $s_1$ and $s_2$. In Fig.~\ref{fig:bc1-zoom} we show the energy optimization of several solution families. Notice that for $\mu_3$, $k_3$, and $c_3$, the optimization curve does not have a robustly defined minimum. Instead one observes a distinct serrated shape, with minimum located very close to the divergence. The reason for the sudden change in energy lies in the nonlinearity of the corresponding equations. Once a certain critical value of the parameter is reached, the solution changes its nature and begins to develop the ``hump'' similar to the one shown in Fig.~\ref{fig:u2straight}. This results in a drastic increase in the variational energy.
The effect is more pronounced for large system sizes. An alternative way to look at the results for $\mu_3$, $k_3$, and $c_3$ is that, while the zero-parameters Eq.~(\ref{eq:original-chopped}) produces rather satisfactory variational energy, it is hard to parametrize. 

Only three equations optimized to an expression significantly different from (\ref{eq:original-chopped}). Eq.~(\ref{eq:original-17}) had minimum with $c_2=0.68$; however, the variational energy was not a smooth function of $c_2$ at the minimum, as can be seen in Fig.~\ref{fig:bc1-zoom}. Both Eqs.~(\ref{eq:original-9}) and~(\ref{eq:original-18}) with parameter $\delta$ and $c_3$ optimized with parameters away from unity. While the serrated shape is still present, it is not as sharp, and both $\delta$ and $c_3$ have minima sufficiently away from the divergence. These are therefore the two equations that are most promising for our purpose. Notice also how broad is the minimum for $\delta$.

\subsection{Dependence on cutoff and the two-parameter BC2 wavefunctions}

\begin{figure*}
  \begin{centering}
  \begin{minipage}[b]{0.49\textwidth}
    \includegraphics[width=\textwidth]{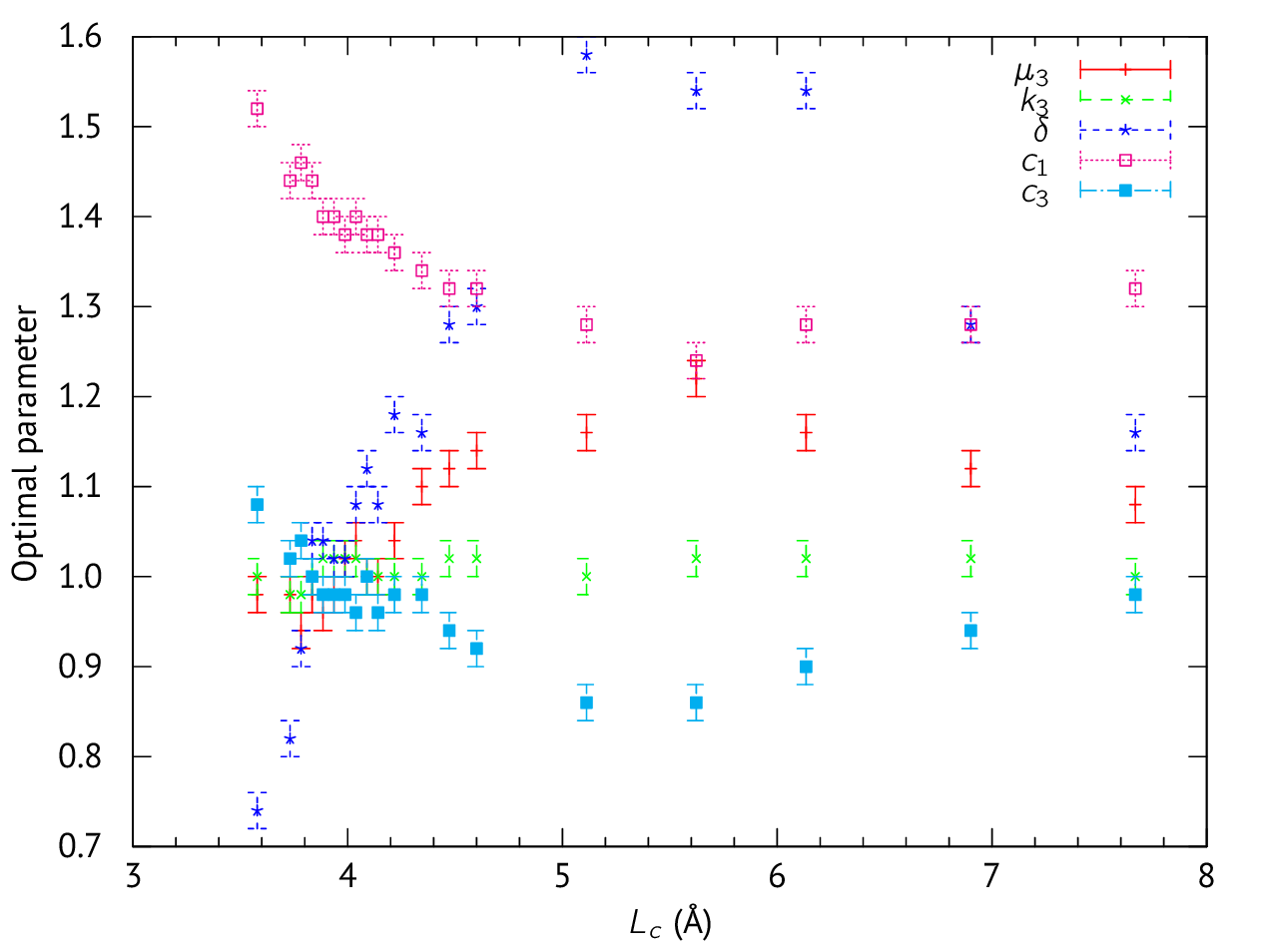}
  \end{minipage} 
  \begin{minipage}[b]{0.49\textwidth}
    \includegraphics[width=\textwidth]{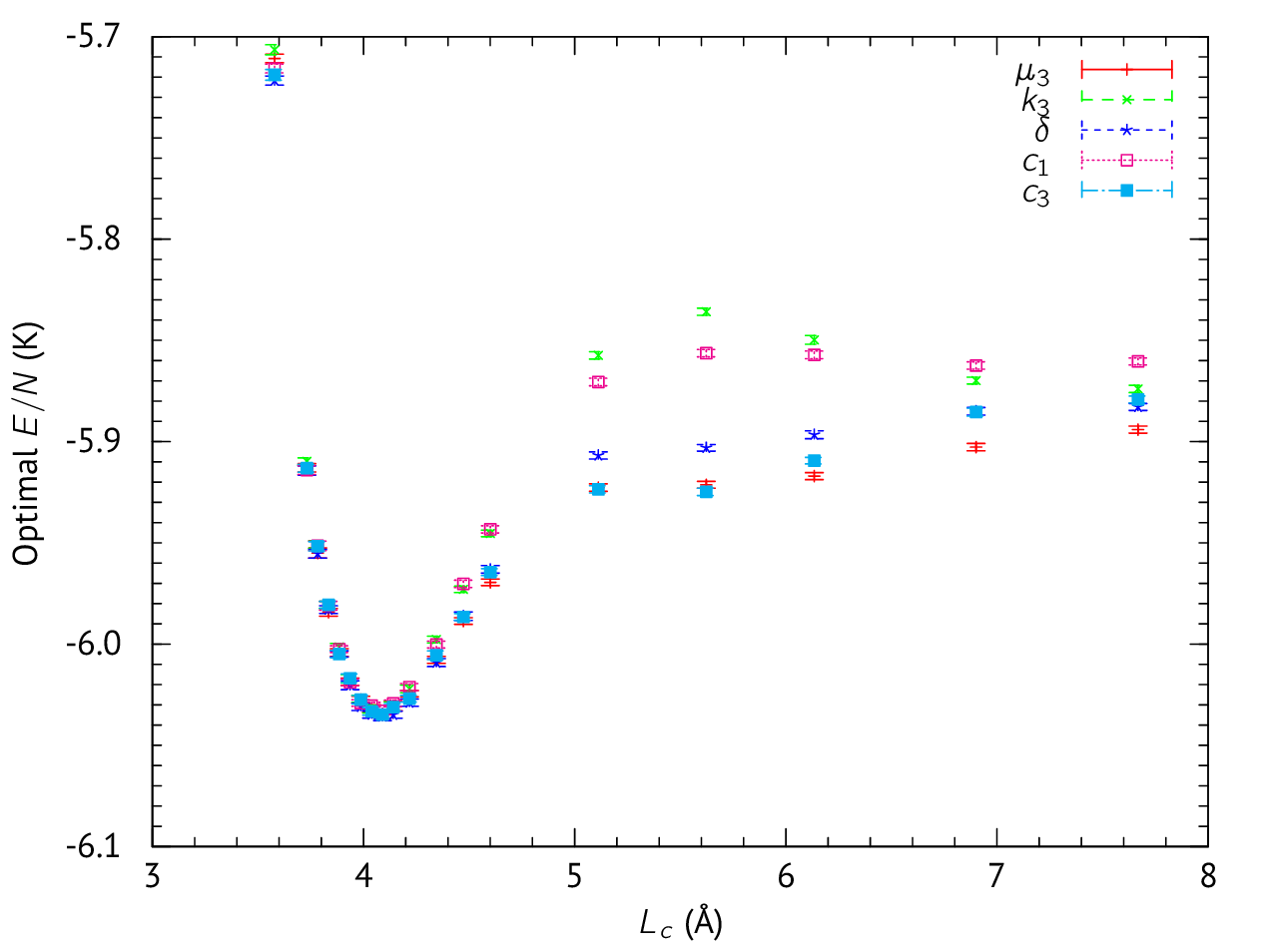}
  \end{minipage}
  \end{centering}
  \caption{\label{fig:bc2-optimization} %
Optimization of two-parameter \bcdos\ pair factors. \emph{Left}: Energy-optimal value of the pair equation modification variable as a function of the cutoff length. \emph{Right:} Optimized energy depending the on the cutoff length.%
  }
\end{figure*}

Here we investigate the two-parameter \bcdos\ pair factors described by Eq.~(\ref{eq:bc2}). The variational parameters are the cutoff length $L_c$ and one parameter for the modified scattering equation. The size of the system, and the cutoff for the potential energy, remain the same as above. The pair equations which failed to produce a satisfactory solution for the largest cutoff (see Table~\ref{tab:results-helium-bc1}) are not considered. This leaves parameters $\mu_3$, $k_3$, $\delta$, $c_1$, and $c_3$. The behavior of these solution families with respect to the cutoff length $L_c$ are shown in Fig.~\ref{fig:bc2-optimization}. The optimal value of parameters $\mu_3$, $k_3$, and $c_3$ remain close to unity, thus close to the divergence seen in Fig.~(\ref{fig:bc1-zoom}). Parameter $c_1$ changes monotonically and only by a small amount, unlike the parameter $\delta$. 

\begin{figure*}
  \begin{centering}
  \begin{minipage}[b]{0.49\textwidth}
    \includegraphics[width=\textwidth]{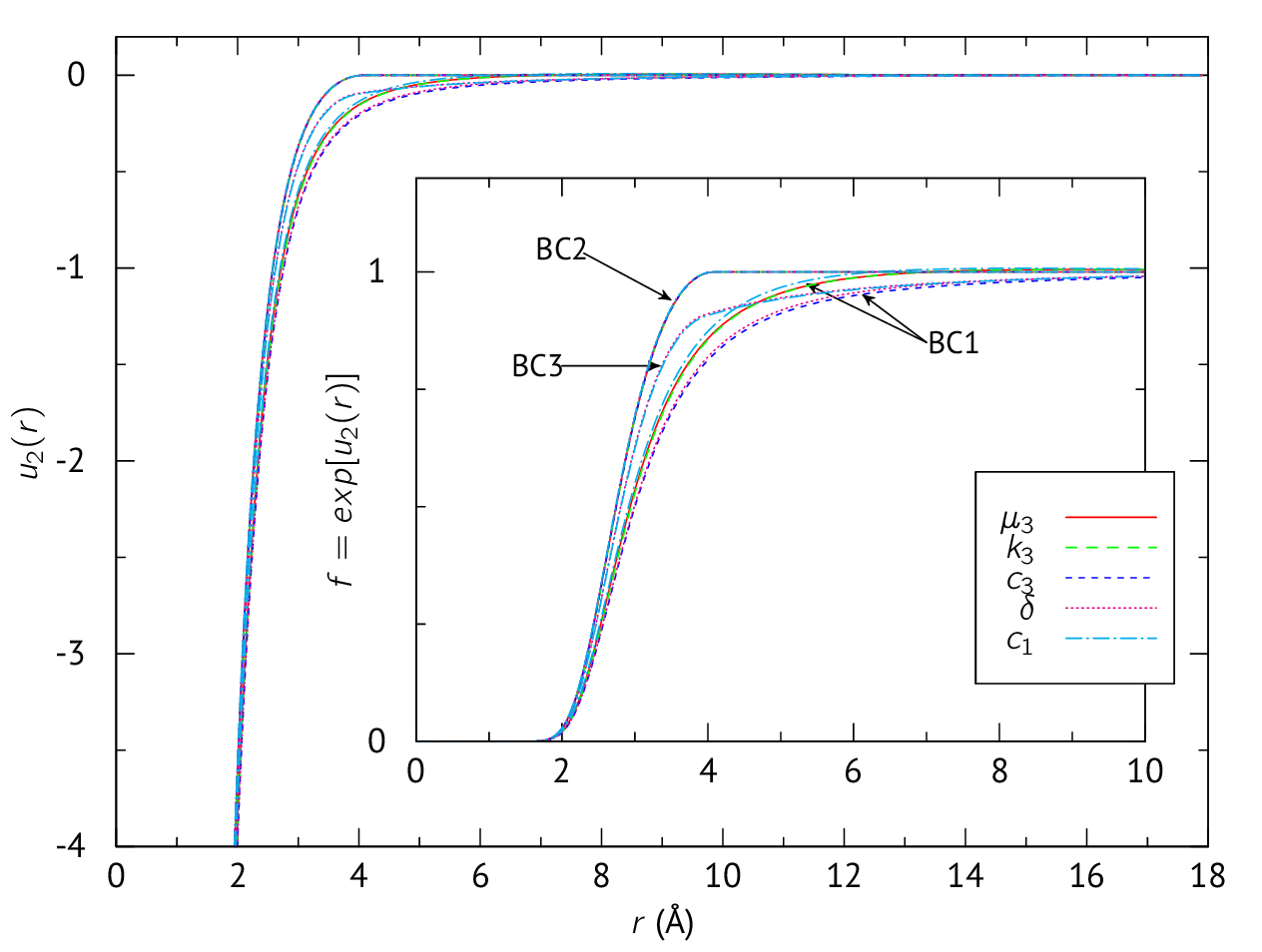}
  \end{minipage} 
  \begin{minipage}[b]{0.49\textwidth}
    \includegraphics[width=\textwidth]{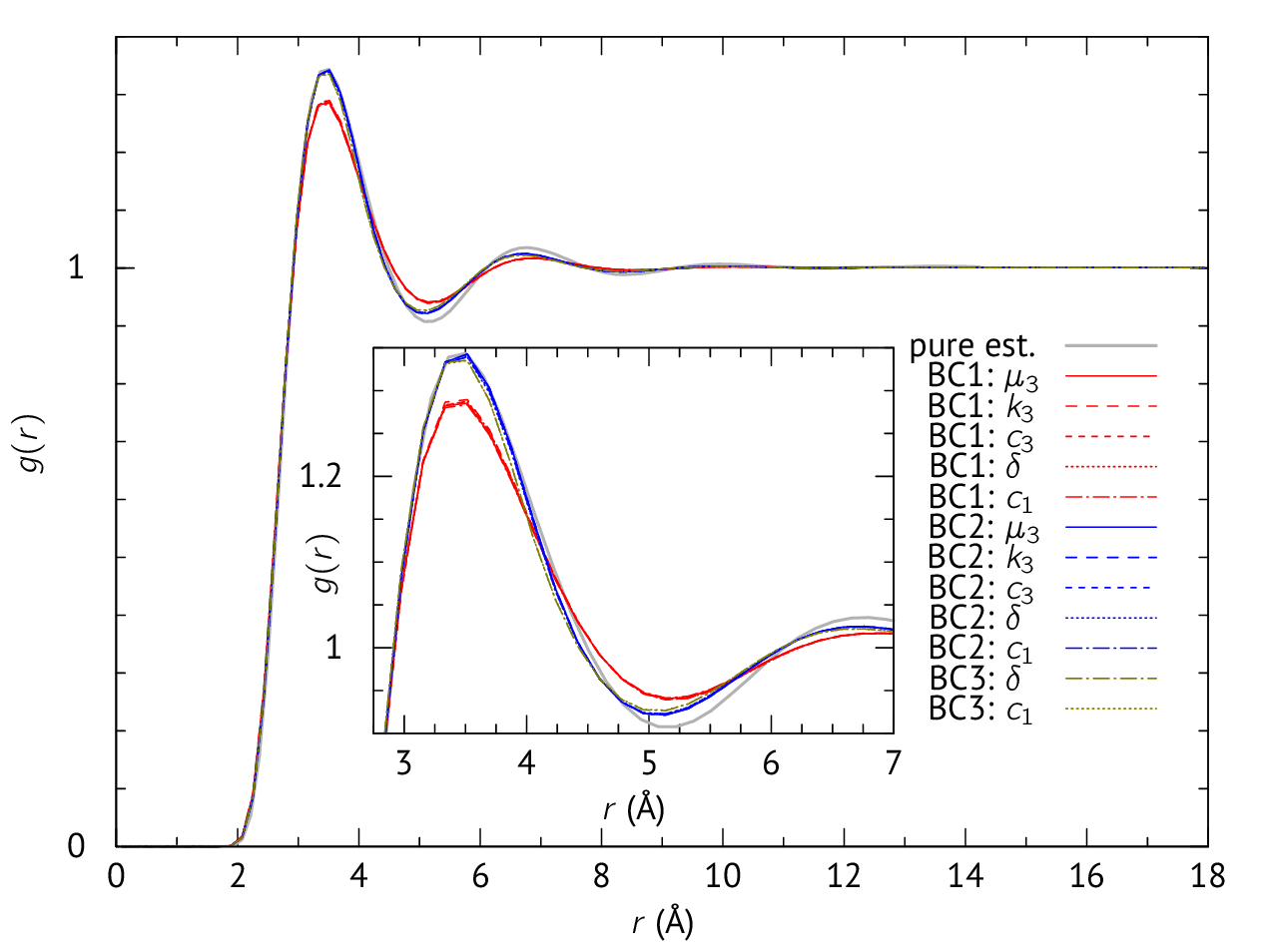}
  \end{minipage}
  \end{centering}
  \caption{\label{fig:pair-functions} %
The pair factors (\emph{left}) and the resulting pair distributions (\emph{right}).
For the one-parameter BC1 function, the optimal pair factors collapse on two distinct and nearly equal solutions. However, the resulting pair distributions are identical. 
For the variable cutoffs of BC2 and BC3, the optimal cutoff is rather small and the solutions nearly collapse onto a single function.
The right panel shows the pure DMC estimator for the pair distribution function shown in  Ref.~\onlinecite{CoordinatedWfn2015}.
It can be seen that both BC2 and BC3 functions reproduce the first peak nearly exactly.
  }
\end{figure*}

Figure~\ref{fig:bc2-optimization} shows optimal energy for each value of the cutoff. All pair factors exhibit an energy minimum at $L_c=4.1$~\AA. This is convenient, as even the smallest system sizes would be able to accommodate such a small cutoff. It is not surprising that $\mu_3$, $k_3$, and $c_3$ coincide at the minimum. After all, they all optimize close to the Eq.~(\ref{eq:original-chopped}). It is much more surprising that $\delta$ and $c_1$, valued away from unity, both produce the same optimal energy value. The resulting pair functions are, in fact, also nearly identical, as shown in Fig.~\ref{fig:pair-functions}. The pair distribution function computed with optimal values of the \bcdos\ wavefunction is shown in the left panel of Fig.~\ref{fig:pair-functions}. The first correlation peak in the pair distribution is replicated almost exactly.
%$L_c=1.6\sigma$. = %e 

\subsection{Dependence on speed of sound and the three-parameter wavefunctions}

Here we consider the phonon tail as given by (\ref{eq:bc3}). The variational parameters are the \new{parameter $c$ related to the speed of sound}, the point $L_c$ at which the solution is joined with the phonon tail, and the value of the parameter of one of the modified pair equations. Given the results above, we only consider the solution families with parameter $\delta$ and $c_1$. It is known that the zero-point phonons do not contribute significantly to the ground state energy of liquid helium. Moreover, we are not using the cutoff wavevector for the long-range tail and the tail must be symmetrized at the periodicity cutoff $L_b$. One thus should expect a limited gain from addition of this extra term. Indeed, we observe a nearly negligible improvement in energy. However, the pair factors could be optimized for a very broad range of values of \new{the parameter} $c$. That is, it is possible to attach additional tail behavior to our functions, while preserving or even improving the variational energy. Following Ref.~%
\onlinecite{ReattoChester1967-PhononsAndThePropertiesOfABoseSystem} and assuming the exponential cutoff for phonon wavevector, the speed of sound $v_s$ is given by
\begin{equation}
v_s\hbar=\frac{\hbar^2}{2m} 4\pi^2 \rho c.
\end{equation} 
We find a very shallow minimum at $v_s\approx 120$~m/s, a two-fold deviation from the actual speed of sound in liquid helium, which reaches 240~m/s at lowest wavenumbers
\cite{Wilhelm1938-TheVelocityOfSoundInLiquidHelium,DonnellyHeliumData}. However, given the weak dependence of energy on the speed of sound parameter $c$, the absence of the cutoff parameter $k_c$ in our function, the dependence of the speed of sound $v_s$ on the wavevector $k$, and the fact that optimal joining point $L_b$ is in fact quite small and smaller that expected values of wavevector cutoff $k_c^{-1}$, this discrepancy is not very surprising.
Optimized values of the parameters are shown in Table~\ref{tab:results-helium-bc1}.

\section{Comparison with past results}

Comparison with published variational results for helium is somewhat obscured by the fact that a number of models for the pair potential has been used throughout the years. Changing the potential can change the optimized energy by as much as 0.1~K in energy per particle. This is illustrated in Table~\ref{tab:comparison} for the McMillan one-parameter function. The original result obtained by McMillan \cite{McMillan} with the Lennard-Jones potential for a rather small system was 0.07~K higher than the same function when used with Aziz HFDHE2 potential by Vitiello et al.~\cite{Vitiello1990-ShadowWaveFunctionVariationalCalculationsOfCrystallineAndLiquidPhasesOfHe4}. Overall, it appears that the explicit forms of Jastrow functions, constructed in the best possible manner, cannot surpass by a considerable margin the barrier of six degrees per atom. However, we find that our two-parameter Jastrow function produces energies down to $-6.04$~K. This is a considerable advantage. Only a small part of it can be attributed to the numerical bias or the difference in the potential model.

\begin{table*}
\begin{ruledtabular}
\begin{center}
\caption{ \label{tab:comparison}
Past variational results for helium with Jastrow functions.
Pair potential models are labeled as LJ for Lennard-Jones with deBoer parametebers (Ref.~\onlinecite{deBoerMichels1939}); 
HFDHE2 and \mbox{HFD-B(HE)} for the corresponding Aziz potentials from Refs.~\onlinecite{Aziz1979} and \onlinecite{AzizII}.
PPA refers to the paired phonon analysis of Ref.~\onlinecite{Campbell1978-FunctionalOptimizationOfTheJastrowWaveFunctionForLiquid4He}.
}
\begin{tabular}{clllcccc} 
Ref.                                 && Pair factor parametrization                                & $E/N$ (K)  &  Pair-potential\\ \hline
\onlinecite{McMillan} 
& (McMillan 1965)                    & $b$ in $u_2=-1/2 (b/r)^5$                  & $-5.65$       &  LJ \\
\onlinecite{Vitiello1990-ShadowWaveFunctionVariationalCalculationsOfCrystallineAndLiquidPhasesOfHe4}
& (Vitiello et al. 1990)             & $b$                                        & $-5.717(21)$  &  HFDHE2 \\
\onlinecite{Boronat1994}
& (Boronat et al. 1994)              & $b$                                        & $-5.683(14)$  &  HFD-B(HE) \\
\onlinecite{Reatto1974-HowGoodCanJastrowWavefunctionsBeForLiquidHeliumFour}
& (Michelis et al. 1974)             & $b, r_0, A, \lambda, \Lambda, C, d, D$     & $-5.96(20)$   &  LJ \\
\onlinecite{Chester1981-ModernPotentialsAndThePropertiesOfCondensedHeFour} 
& (Kalos et al. 1981)                & PPA                                        & $-5.87$       &  HFDHE2 \\ 
\onlinecite{Vitiello1992-OptimizationOfHeWaveFunctionsForTheLiquidAndSolidPhases}
& (Vitiello et al. 1992)             & 2B(BS) basis set, $c_1,\cdots,c_{10}$       & $-5.938(28)$  &  HFDHE2 \\
\onlinecite{Moroni1995-EulerMonteCarloCalculationsForLiquidHe4AndHe3}
& (Moroni et al. 1995)               & Linear set, tens of coef.                   & $-6.001(16)$  &  HFDHE2 \\
\end{tabular}
\end{center} 
\end{ruledtabular}
\end{table*}

%$b, r_0, A, \lambda, \Lambda, C, d, D$

\section{Conclusions}

We have revisited the long-posed question of producing satisfactory Jastrow wavefunction based on the corresponding pair scattering equation. 
We proposed and considered in detail a straight-forward method for producing high-quality pair factors based solely on the two-body potential of the system. The method expands on the pair factors used some time ago by Pandharipande. We propose two-parameter pair factors, in which one parameter modifies the pair equation, such as $\delta$ in Eq.~(\ref{eq:original-9}) or $c_1$ in Eq.~(\ref{eq:original-16}), and the second parameter is the solution cutoff. The modified equations preserve the leading-order solution of the pair equation at short distances. Such a pair factor provides excellent variational optimization and reproduces the first peak of the pair-distribution function. It is possible to add known or desired long-range behavior to such pair factors.
It should be concluded that not only it is possible to extract satisfactory pair factors from the pair scattering equation, but that such factors are of high-quality and may be of convenience due the low number of corresponding variational parameters.

\section{Acknowledgments}

I am thankful to Jordi Boronat, Grigory Astrakharchik, Siu A. Chin, and Eckhard Krotscheck for useful discussions. 
The author acknowledges the computer resources and
assistance provided by the Spanish Supercomputing Network
(Red Espa\~{n}ola de Supercomputaci\'on), and by the Barcelona
Supercomputing Center. Hardware and technical expertise was also provided by the Nvidia corporation. The author gratefully acknowledges the Gauss Centre for Supercomputing e.V.\ (www.gauss-centre.eu) for funding this project by providing computing time on the GCS Supercomputer SuperMUC at Leibniz Supercomputing Centre (www.lrz.de).

% %\section*{References}
% \vspace*{2em}
% %\bibliographystyle{plain}
% \bibliography{supersolidity,MyPapers,MCMethods}

% \bibliography{/home/yaroslav/projects/bibs/supersolidity,%
% /home/yaroslav/projects/bibs/MyPapers,%
% /home/yaroslav/projects/bibs/MCMethods%
% }


\end{document}